\documentclass[twocolumn,showpacs,amsmath,amssymb,aps,prl,floats]{revtex4-1}
\usepackage{bm}
\usepackage{graphicx}
\usepackage{times}
\usepackage{dcolumn}
\newcommand{\EQ}{\begin{equation}}
\newcommand{\EN}{\end{equation}}
\newcommand{\EQA}{\begin{eqnarray}}
\newcommand{\ENA}{\end{eqnarray}}

\newcommand{\xx}{\mbox{\boldmath $x$} {}}
\newcommand{\nn}{\mbox{\boldmath $n$} {}}
\newcommand{\kk}{\mbox{\boldmath $k$} {}}
\newcommand{\KK}{\mbox{\boldmath $K$} {}}
\newcommand{\qq}{\mbox{\boldmath $q$} {}}
\newcommand{\bb}{\mbox{\boldmath $b$} {}}

\newcommand{\ssvec}{\mbox{\boldmath $s$} {}}

\newcommand{\kkk}{\hat{\bm{k}}}

\newcommand{\rrr}{\hat{\bm{r}}}

\begin{document}

\title{Cosmic Microwave Background Trispectrum and Primordial Magnetic Field Limits}

\author{Pranjal Trivedi$^{1,2}$}
\email{ptrivedi@physics.du.ac.in} 
\author{T. R. Seshadri$^{1}$}
\author{Kandaswamy Subramanian$^{3}$}

\affiliation{
$^1$Department of Physics and Astrophysics, University of Delhi, Delhi 110007, India. \\
$^{2}$Department of Physics, Sri Venkateswara College, University of Delhi, Delhi 110021, India.\\
${^3}$IUCAA, Post Bag 4, Ganeshkhind, Pune 411 007, India.
}
\date{\today}

\begin{abstract}
Primordial magnetic fields will generate non-Gaussian signals in the cosmic microwave
background (CMB) as magnetic stresses and the temperature anisotropy they induce 
depend quadratically on the magnetic field. 
We compute a new measure of magnetic non-Gaussianity, the CMB trispectrum, on large angular scales, 
sourced via the Sachs-Wolfe effect.
The trispectra induced by magnetic energy density and by magnetic scalar anisotropic stress are found to have
typical magnitudes of approximately a few times $10^{-29}$ and $10^{-19}$, respectively.
Observational limits on CMB non-Gaussianity from WMAP data allow us to conservatively set upper limits 
of a nG, and plausibly sub-nG, on the present value of the primordial cosmic magnetic field.
This represents the tightest limit so far on the strength of primordial magnetic fields, on Mpc scales, 
and is better than limits from the CMB bispectrum and all modes in the CMB power spectrum.
Thus, the CMB trispectrum is a new and more sensitive probe of primordial magnetic fields on large scales.

\end{abstract}

\maketitle

Magnetic fields are ubiquitous in the Universe from planets and stars to
galaxies and galaxy clusters \cite{B_obs_gal_clusters,BS05_KZ08_W02}, yet the origin and evolution of 
large-scale magnetic fields remains a puzzle. 
A popular paradigm is that magnetic fields in collapsed structures
could arise from dynamo amplification of seed magnetic fields \cite{BS05_KZ08_W02}.
The seed field could in turn be generated in astrophysical batteries \cite{AstroBatteries}
or due to processes in the early universe \cite{Inflation_B,PhaseTrans_B}.
Indeed recent $\gamma$-ray observations claim to find a lower limit
to an all-pervasive intergalactic magnetic field that fills most of 
the cosmic volume \cite{neronov10},
which would perhaps favor a primordial origin.
A primordial magnetic field can be generated at 
inflation \cite{Inflation_B}, or 
arise out of other phase transitions in the early Universe \cite{PhaseTrans_B}. 
As yet there is no compelling mechanism which produces strong
coherent primordial fields.
Equally, the dynamo paradigm is not without its own challenges
in producing sufficiently coherent fields and sufficiently rapidly \cite{BS05_KZ08_W02}.
Therefore, it is useful to keep open the possibility that
primordial magnetic fields originating in the early universe play a
crucial role in explaining the observed cosmic magnetism.

In this context it is important to investigate every observable 
signature of the putative primordial magnetic fields.
Constraints on large-scale primordial magnetic fields have already
been derived using the cosmic microwave background (CMB) power spectrum 
\cite{yam10_pao10_SL10,CMB_B_reviews} and Faraday rotation \cite{FR_CMB_B}. 
However, the effects of a magnetic field on the CMB 
are relatively more prominent in its non-Gaussian correlations.
This is because magnetic fields induce non-Gaussian signals at lowest order 
as the magnetic energy density and stress are quadratic in the field. 
On the other hand, the standard inflationary perturbations, 
dominated by their linear component, can source non-Gaussian correlations 
only with higher-order perturbations and thus necessarily produce a small amplitude of 
CMB non-Gaussianity (cf. \cite{NG_inflation,komatsu_spergel01}).
Primordial magnetic fields can induce appreciable CMB non-Gaussianity when considering the bispectrum 
\cite{SS09,caprini09_cai10}. 
Our previous calculation of the magnetic CMB bispectrum sourced by 
scalar anisotropic stress led to a 
$\sim 2$ nG upper limit on the primordial magnetic field's amplitude on Mpc scales \cite{TSS10}.
However, higher-order measures of non-Gaussianity remain unexplored and
as we show here, could be very useful to set further constraints 
on primordial magnetic fields.

In this letter, we present the first calculation 
of the contribution to the CMB trispectrum 
induced by a primordial magnetic field. In particular, we consider 
the magnetically induced Sachs-Wolfe effect 
sourced by a stochastic primordial magnetic field. 
We show that the trispectrum does significantly better than the bispectrum in constraining 
the large-scale magnetic field via CMB non-Gaussianity, considering both 
magnetic energy density and magnetic scalar anisotropic stress as sources. 
This reveals a new and effective probe to investigate primordial magnetic fields on large scales.

We consider a Gaussian random stochastic magnetic field ${\bf B}$ characterized and completely specified 
by its power spectrum $M(k)$. We further assume the magnetic field to be nonhelical. On galactic and larger 
scales, any velocity induced by Lorentz forces is generally too small to 
appreciably distort the initial magnetic field \cite{jedam98_SB98}.  Hence, the magnetic field 
simply redshifts away as ${\bf B}({\bf x},t)={\bb}_{0}({\bf x})/a^{2}$, where, ${\bb}_{0}$ 
is the magnetic field at the present epoch (i.e. at $z=0$ or $a=1$).
We define ${\bb} ({\kk})$ as the Fourier transform of the magnetic field ${\bb}_0 ({\xx})$. 
The magnetic power spectrum is defined by the relation 
$\langle b_{i}({\kk})b^*_{j}({\qq})\rangle=(2\pi)^3 \delta({{\kk}-{\qq}})P_{ij}({\kk})M(k)$, 
where $P_{ij}({\kk}) = (\delta_{ij} - k_ik_j/k^2)$ is the projection operator ensuring 
${\bf \nabla}\cdot{\bb_0} =0$. 
This leads to $\langle {\bb}_{0}^{2} \rangle=2\int (dk/k)\Delta _{b}^{2}(k)$, 
where $\Delta _{b}^{2}(k)=k^{3}M(k)/(2\pi ^{2})$ 
is the power per logarithmic interval in $k$ space present in the stochastic magnetic field.
We assume a power-law magnetic 
power spectrum,
$M(k)=Ak^{n}$ that has a cutoff at $k=k_{c}$,
where $k_{c}$ is the Alfv\'{e}n-wave damping length scale
\cite{jedam98_SB98}. We fix $A$ by setting the variance of the magnetic field to be $B_0$, smoothed using a sharp $k$-space
filter, over a ``galactic`` scale $k_G=1h$ Mpc$^{-1}$. This gives (for $n \gtrsim -3$ and for $k<k_c$)
\EQ
\Delta _{b}^{2}(k)= \frac{k^3M(k)}{2\pi^2}
=\frac{B_0^2}{2}(n+3)\left(\frac{k}{k_{G}}\right)^{3+n}.
\EN
The spectral index $n$ is restricted to values close to and above -3, i.e., an inflation-generated field, 
as causal generation mechanisms can only produce much bluer spectra \cite{durrer_caprini03}. 
Further, blue spectral indices 
are strongly disfavored by many observations like the CMB power spectra \cite{yam10_pao10_SL10}. 
We choose to split the contributions to the CMB trispectrum into that sourced by magnetic energy density $\Omega_B$ 
and by scalar anisotropic stress $\Pi_B$ rather than the compensated and 
passive magnetic perturbation modes of Ref. \cite{SL09}. 
The subdominant compensated mode is a linear combination of $\Omega_B$ and $\Pi_B$ whereas the passive mode 
is the $\Pi_B$ perturbation considered here.

The Sachs-Wolfe type of contribution to the CMB temperature anisotropy induced by 
the energy density of magnetic fields \cite{giovannini07_PMC,pao09_fin08,bonvin10},
can be expressed as
\EQ
\frac{\Delta T}{T}(\nn) =  
{\cal R} ~ \Omega_B(\xx_0 -\nn D^*).
\EN
Here, $\Omega_B({\xx}) = {\bf B}^2({\xx},t)/(8\pi \rho_{\gamma}(t)) 
={\bb}_0^2({\xx})/(8\pi \rho_0)$, 
where $\rho_{\gamma}(t)$ and $\rho_0$ are, respectively, the CMB energy densities at a time $t$ and 
at the present epoch. In the same manner as the usual Sachs-Wolfe effect, the $\Delta T/T$ 
given above is for large angular scales.  
For numerical estimates we use the most recent estimate of Bonvin and Caprini (Eq. 6.12 of \cite{bonvin10}) 
expressed according to our definitions as ${\cal R} = -0.2 R_{\gamma} / 3 \sim -0.04 $ where
$R_{\gamma} \sim 0.6$ is the fractional 
contribution of radiation energy density towards the total energy density of the relativistic 
component. The unit vector ${\bf n}$ is along the direction of observation from the observer 
at position $\xx_0$ and $D^*$ is the 
(comoving angular diameter) distance to the surface of last scattering. We have assumed instantaneous 
recombination which is a good approximation for large angular scales.

The temperature fluctuations of the CMB can be expanded in terms of spherical harmonics to give 
$\Delta T(\nn)/T = \sum_{l m} a_{lm} Y_{lm}(\nn)$,
where
\EQ
a_{lm}= \frac{4 \pi}{i^l}\int \frac{d^3 k}{(2\pi)^3} ~
{\cal R} ~ \Omega_B(\kk) ~ j_{_{l}}(kD^*) ~ Y^*_{lm}(\kkk). 
\label{alm}
\EN
Here, $\Omega_B(\kk)$ is the Fourier transform of $\Omega_B({\xx})$. 
Since $\Omega_B({\xx})$ is quadratic in ${\bb}_0({\xx})$, we have a convolution 
$\Omega_B(\kk) = \left({1}/{(2\pi)^{3}}\right)\int d^3s ~ b_i(\kk+\ssvec) b^*_i(\ssvec)/(8\pi \rho_0)$.
The trispectrum $T^{m_{_1}\!m_{_2}\!m_{_3}\!m_{_4}}_{\,\,\,\,l_{_1}\,\;l_{_2}\,\;l_{_3}\,\;l_{_4}}$, 
or the four-point correlation function of the CMB temperature anisotropy in harmonic space, in terms of the ${a_{lm}}$'s, is
$T^{m_{_1}\!m_{_2}\!m_{_3}\!m_{_4}}_{\,\,\,\,l_{_1}\,\;l_{_2}\,\;l_{_3}\,\;l_{_4}}
=\,\langle a_{{l_1}{m_1}}a_{{l_2}{m_2}}a_{{l_3}{m_3}}a_{{l_4}{m_4}}\rangle$.
From Eq. (\ref{alm}) we can express $T^{m_{_1}\!m_{_2}\!m_{_3}\!m_{_4}}_{\,\,\,\,l_{_1}\,\;l_{_2}\,\;l_{_3}\,\;l_{_4}}$ as
\EQ
\!\!T^{m_{_1}\!m_{_2}\!m_{_3}\!m_{_4}}_{\,\,\,\,l_{_1}\,\;l_{_2}\,\;l_{_3}\,\;l_{_4}}
\!= \! \left(\!\frac{\cal R}{2\pi^2}\!\right)^{\!\!4}\!\!\int 
\!\left[\prod_{i=1}^4
\frac{d^3 k_i}{i^{l_i}} j_{_{l_i}}\!(k_{_i}D^*)Y^*_{l_im_i}\!(\hat{\kk}_{_i})\right]
\!\zeta_{_{1234}}
\label{trispec}
\EN
with 
$\zeta_{_{1234}}=\, \langle \Omega_B(\kk_{_1})\Omega_B(\kk_{_2})
\Omega_B(\kk_{_3})\Omega_B(\kk_{_4})\rangle$.
The four-point correlation function of $\Omega_B(\kk)$ involves an eight-point 
correlation function of the fields. Using Wick's theorem, for Gaussian magnetic fields, 
we can express the magnetic eight-point correlation as a sum of 105 terms involving the magnetic 
two-point correlation. Neglecting the 45 terms proportional to $\delta(\kk)$ that vanish and 
the 12 terms proportional to $\delta(\kk_i + \kk_j)$
that represent the unconnected part of the four-point correlation, we are left with 
48 terms. A long calculation involving the relevant projection operators gives
$\zeta_{_{1234}} = 
\delta(\kk_1 + \kk_2 + \kk_3 + \kk_4) ~ \psi_{_{1234}}$, where $\psi_{_{1234}}$ is a mode-coupling
integral over a variable $\ssvec$ and also involves angular terms.
The full expression for $\psi_{_{1234}}$ will be presented in 
our detailed paper \cite{TSS11_passive_trispec}.
For simplicity we evaluate the mode-coupling integral $\psi_{_{1234}}$ in two cases: 
(I) considering only $\ssvec$-independent angular terms for all equal-sided configurations and 
(II) taking all angular terms for the collinear configuration. 
Considering $\ssvec$-independent terms only for a general configuration, we find
$\psi_{_{1234}} = {-8}/{(8\pi\rho_0)^4} ~ {\cal I}$
where
\EQA
{\cal I} &=& \int d^3 s \; M(s) \; M(\vert \kk_1 + \ssvec \vert) \times \nonumber \\ 
  \Big[ &M& (\left| \kk_1 + \kk_3 +\ssvec \right|) \Big( M(\vert \kk_2  -  \ssvec \vert) 
                                                + M(\vert \kk_4  -  \ssvec \vert) \Big) \nonumber \\
+ &M&(\left| \kk_1 + \kk_2 +\ssvec \right|) \Big( M(\vert \kk_3  -  \ssvec \vert)  
                                                + M(\vert \kk_4  -  \ssvec \vert) \Big) \nonumber \\
+ &M&(\left| \kk_1 + \kk_4 +\ssvec \right|) \Big( M(\vert \kk_2  -  \ssvec \vert) 
                                                + M(\vert \kk_3  -  \ssvec \vert) \Big) \Big] \nonumber \\
&=& \; \; {\cal I}_{\rm (1)} +  {\cal I}_{\rm (2)} + {\cal I}_{\rm (3)} + {\cal I}_{\rm (4)} 
+ {\cal I}_{\rm (5)} + {\cal I}_{\rm (6)}.
\label{calI_I}
\ENA
We perform the mode-coupling integral using the technique and approximations 
discussed in \cite{TSS10,trsks01_mack02_SSB03}, while adopting the mean (zero) value of $\kkk_1 \cdot \kkk_3$, to find
\EQ
{\cal I}_{\rm (1)} \simeq 4 \pi A^4 ~ k_1^{2n+3} ~ k_2^n ~ k_3^n \left[ \frac{2^{n/2}}{n+3} - \frac{1}{4n+3} \right].
\label{Ii}
\EN
The value of each of the ${\mathcal I}_{(j)}$ integrals for $j={\rm 1} - {\rm 6}$ is the same 
when all the $\vert \kk_i \vert \simeq k$. We perform the $\ssvec$-independent (case I) trispectrum 
evaluation for such equal-sided quadrilateral configurations. Hence, ${\mathcal I}= \sum_{j=(1)}^{(6)} 
{\mathcal I}_j 
= 6 \,\, {\mathcal I}_{\rm (1)}$, and 
we obtain
\EQA
&\!&\zeta_{_{1234}} \; \simeq \;\;\delta(\kk_1 + \kk_2 + \kk_3 + \kk_4) \times \nonumber \\
&&\!\! \!\! \frac{-8 \, (24\pi)\, A^4 \,k_1^{2n+3} \,k_2^n \,k_3^n }{(8\pi\rho_0)^4}\!\! \left[ \frac{(2^{n/2})(4n+3)-(n+3)}{(4n+3)(n+3)} \right]\!.
\label{zeta_I}
\ENA
Inserting this into Eq. (\ref{trispec}) for the trispectrum and following the approach of 
\cite{OkamotoHu02_KogoKomatsu06}, we decompose our delta function as
$ \delta(\kk_1 + \kk_2 + \kk_3 + \kk_4) = \int d^3 K \delta(\kk_1 + \kk_2 + \KK)\delta(\kk_3 + \kk_4 - \KK)$.
Using the integral form of the delta functions
and the spherical wave expansion
we perform the integrations over the angular parts of $(\kk_1,\kk_2,\kk_3,\kk_4,\KK)$, with algebra
similar to \cite{FS07,SS09,TSS10}, to give
\EQA
\!\!&&T^{m_{_1}\!m_{_2}\!m_{_3}\!m_{_4}}_{\,\,\,\,l_{_1}\,\;l_{_2}\,\;l_{_3}\,\;l_{_4}}
\!\simeq \left[  \frac{(-768) \left( A {\cal R}\right)^4} {\pi^7 \left( 8\pi \rho_0 \right)^4 } \right] \!\! \left\lbrace \frac{(2^{n/2})(4n+3)-(n+3)}{(4n+3)(n+3)} \right\rbrace \nonumber \\
&&  \! \times \!\! \int 
\!\left[\prod_{i=1}^4
d k_i \, k_i^2 \, j_{_{l_i}}\!(k_{_i}D^*) \, j_{_{l_i}}\!(k_{_i} \bar{r}_{_i}) \right]
\!k_1^{2n+3} \left(k_2 k_3 \right)^n \nonumber \\
&& \! \times  \sum_{LM} (-1)^{L-M} \int dK  K^2  j_{_{L}}(Kr_1) \, j_{_{L}}(-Kr_2) \nonumber \\
&& \! \times \!\! \int \!\left[\prod_{i=1}^2
d^3 r_i \, Y_{l_{2i-1} m_{2i-1}}\!(\rrr_i) \, Y_{l_{2i} m_{2i}}\!(\rrr_i) \, Y_{L \; (-1)^{i+1}\!M}(\rrr_i) \right] 
\label{trispec_after_k_angular}
\ENA
with $\bar{r}_{_i}$ equal to  $r_1$ for $i=1,2$ and $r_2$ for $i=3,4$.
The approximations involved (with respect to angular terms) in the $\kkk_i$ angular integrals 
can be made more precise by going to the flat-sky limit (elaborated in our detailed paper 
\cite{TSS11_passive_trispec}).
Here the $K$ integral gives $\delta(r_1 - r_2)\left({\pi}/{2 r_1^2}\right)$ via the spherical Bessel 
function closure relation. This delta function enables us to perform the $r_2$ integral trivially, 
then $r_1$ replaces $r_2$ in the arguments of $j_{_{l_3}}$ and $j_{_{l_4}}$. The angular 
$\rrr_1$ and $\rrr_2$ integrals may be expressed as [e.g. Eq. 5.9.1 (5) of \cite{Varshalovich}]
\EQA
&& \int d\Omega_{\rrr_1} Y_{l_1m_1}(\rrr_1) Y_{l_2m_2}(\rrr_1) Y_{LM}(\rrr_1) = \nonumber \\
&& \!  \sqrt{\frac{(2l_1+1)(2l_2+1)(2L+1)}{4\pi}} 
\! \! \begin{pmatrix} l_1 & l_2 & L\\ 0 & 0 & 0 \end{pmatrix} 
\! \! \! \begin{pmatrix} l_1 & l_2 & L\\ m_1 & m_2 & M \end{pmatrix} \nonumber \\
&& \equiv h_{l_1 L \, l_2} \begin{pmatrix} l_1 & l_2 & L\\ m_1 & m_2 & M \end{pmatrix},
\label{h_l1_L_l2}
\ENA
where we have defined $h_{l_1 L \, l_2}$ above, along the same lines as \cite{OkamotoHu02_KogoKomatsu06}.
We use the relation 
$\left( {A}/{8\pi\rho_0} \right)^4 = \left({2}/{3}\right)^4 \left({\pi}/{k_G}\right)^8 
\left({(n+3)}/{k_G^{n+1}}\right)^4 {V_A}^8$, 
where the Alfv\'en velocity $V_A$, in the radiation dominated era, is defined as 
$V_{A}=B_{0}/\left( 16\pi \rho _{0}/3 \right)^{1/2}\approx 3.8\times 10^{-4}\, B_{-9}$  
\cite{jedam98_SB98}
, with $B_{-9}\ \equiv (B_{0}/10^{-9}{\rm G})$.
From the definition of the rotationally invariant angle-averaged trispectrum \cite{Hu01}
\EQA
\!\!T^{m_{_1}m_{_2}m_{_3}m_{_4}}_{l_{_1}l_{_2}l_{_3}l_{_4}}
&=& \sum_{LM} (-1)^{-M} \begin{pmatrix} l_1 & l_2 & L\\ m_1 & m_2 & -M \end{pmatrix} \nonumber \\
&& \times \begin{pmatrix} l_3 & l_4 & L\\ m_3 & m_4 & M \end{pmatrix} T^{l_{_1}l_{_2}}_{l_{_3}l_{_4}} (L)
\label{angle_avg_trispec},
\ENA
we separate out the reduced trispectrum $T^{l_{_1}l_{_2}}_{l_{_3}l_{_4}} (L)$ (called the 
angular averaged trispectrum in \cite{Hu01}) from the full trispectrum.
We again use the spherical Bessel function closure relation to perform the $k_4$ integral that yields 
$\delta(r_1-D^*)\left({\pi}/{2r_1^2}\right)$.  This facilitates the $r_1$ integral that results in 
$r_1 \rightarrow D^*$ in the arguments of $j_{_{l_1}}$, $j_{_{l_2}}$ and $j_{_{l_3}}$. 
The $k_1$, $k_2$ and $k_3$ integrals containing a product of a power-law and $ j_{_l}^2$ 
can be evaluated in terms of Gamma functions (e.g., Eq. 6.574.2 of \cite{Gradshteyn6}). 
For a scale-invariant magnetic index $n \to -3$, we get
\EQA
\left[ T^{l_{_1}l_{_2}}_{l_{_3}l_{_4}} (L) \right]_{\Omega}
\simeq && \,
-5.8 \times 10^{-29} 
\left(\frac{n+3}{0.2}\right)^3 \left(\frac{B_{-9}}{3}\right)^8 \nonumber \\
 && \times \frac{ h_{l_1 L \, l_2} \, h_{l_3 L \, l_4}}{l_1(l_1+1) l_2(l_2+1) l_3(l_3+1)}.
\label{trispec_energy_density}
\ENA
This gives us the amplitude of the magnetic CMB trispectrum sourced by the 
energy density $\Omega_B$ of a primordial magnetic field.
A factor of $1/(D^* k_G)^{4(n+3)}$ also 
appears which approaches unity for the case $n\! \to\! -3$ of a scale-invariant magnetic field index. We evaluate the
magnetic trispectrum for a near scale-invariant index $n\!=\!-2.8$, for which this factor is  
$\!\sim\! 1/1500$. It turns out that this factor is almost entirely cancelled by the increase in value of the $k$ integrals
when evaluated for $n\!=\!-2.8$ rather than $n\!=\!-3$ \cite{TSS11_passive_trispec}. 

We now compare our magnetic trispectrum with the 
Sachs-Wolfe contribution to the standard CMB trispectrum 
sourced by nonlinear terms in the inflationary perturbations \cite{OkamotoHu02_KogoKomatsu06,Regan10}.
More specifically, in the Sachs-Wolfe limit, the dominant term of
Eq. (64) of Ref. \cite{smidt10} becomes
\EQA
 T^{l_{_1}l_{_2}}_{l_{_3}l_{_4}} (L)  
&\approx& 25 \, \tau_{NL}  \, C_{l_2}^{SW} \, C_{l_4}^{SW} \,  C_{L}^{SW} \, h_{l_1 L \, l_2} \, h_{l_3 L \, l_4} \quad \nonumber \\
&&\! \! \!\! \! \!\! \! \!\! \! \!\! \! \!\! \! \!\! \! \!\! \! \!\! \! \!\!\! \! \!
\approx 5.4 \times 10^{-27} \tau_{NL} \frac{ h_{l_1 L \, l_2} \, h_{l_3 L \, l_4}}{l_1(l_1+1) l_2(l_2+1) l_3(l_3+1)} \, q. 
\label{trispectrum_fNL}
\ENA
Here $\tau_{NL}$ and $f_{NL}$ (below) are standard non-Gaussianity parameters and we adopt the standard estimate for the 
Sachs-Wolfe contribution $C_{l}^{SW}\!$ \cite{OkamotoHu02_KogoKomatsu06}. The factor 
$q$ which is equal to $\!\left[{l_1(l_1+1) l_3(l_3+1)}\right]\!/\!\left[{l_4(l_4+1) L(L+1)} \right]$ 
is of order unity for many configurations. 
Equation (\ref{trispectrum_fNL}) is of the same form as Eq. (\ref{trispec_energy_density}) 
for the magnetic field-induced trispectrum.
We use the negative-sided limit on $\tau_{NL}$ derived from searching for the CMB trispectrum signal in the WMAP5 data \cite{smidt10}, 
$\tau_{NL} > -6000$. 
Magnetic field limits are obtained by taking the one-eighth power of the appropriate ratio of trispectra, 
which gives 
$B_0 \lesssim 16$ nG, 
at a scale of $k_G = 1 h$ Mpc$^{-1}$ for a magnetic spectral index of $n=-2.8$. 
This limit is approximately 2 times stronger than the  
$B_0 \lesssim 30 \text{ nG}$ upper limit for the 
magnetic energy density bispectrum \cite{SS09} 
(taking into account the recent estimate of ${\cal R}$ \cite{bonvin10}),
for the same scale and magnetic index. 

We now calculate the trispectrum for the collinear configuration [case II]. The full mode-coupling 
integral $\psi_{_{1234}}$ \cite{TSS11_passive_trispec} is evaluated over all angular terms for the 
equal-sided collinear configuration $\kk_1 \simeq \kk_2 \simeq -\kk_3 \simeq -\kk_4$.
The four-point correlation of magnetic energy density for the collinear configuration is found to be
\EQA
&\zeta_{_{1234}}& \; \simeq \;\;\delta(\kk_1 + \kk_2 + \kk_3 + \kk_4) \times \nonumber \\
&\!&\!\!\!\!\!\!\!\!\!\!\!\!\!\!\!\!\!\!\frac{8 \, (4\pi)\, A^4 \,k_1^{2n+3} k_2^n k_3^n }{(8\pi\rho_0)^4}
\!\!\left[ \frac{\frac{8}{3}(2^{n/2})(4n+3)-(12)(n+3)}{(4n+3)(n+3)} \right]\!\!.
\label{zeta_coll}
\ENA
Using $n\!=\!-2.8$, we compare the collinear configuration four-point correlation $\zeta$, 
including all angular terms, to $\zeta$ for case I [Eq. \ref{zeta_I}] that included 
only $\ssvec$-independent terms. The collinear $\zeta$ is similar in magnitude but of positive sign
and one then expects a trispectrum also of similar magnitude to case I.

In addition to magnetic energy density, the scalar anisotropic stress associated with a primordial magnetic field 
will also act as a separate source for CMB fluctuations - dominantly in the passive mode \cite{SL09}. 
As we saw in our previous work \cite{TSS10}, the magnetic scalar anisotropic stress generates $\sim 10^6$ times 
larger contribution to the CMB bispectrum compared to magnetic energy density. With this motivation and
using the magnetic trispectrum technique, developed above for energy density, we carry out a longer calculation 
for the trispectrum. The temperature anisotropy, sourced via the magnetic Sachs-Wolfe effect by 
magnetic scalar anisotropic stress $\Pi_B$ [defined in Eq. (6) of \cite{TSS10}, see also \cite{SL09,bonvin10}], is 
\EQ
\frac{\Delta T}{T}(\nn) =  
\mathcal{R}_p ~ \Pi_B(\xx_0 -\nn D^*),
\label{Delta_T_stress}
\EN
where $ \mathcal{R}_p 
= \left[ -R_{\gamma}/15 \right] \ln \left( {T_{B}}/{T_{\nu}} \right)$ and 
$T_B$ and $T_{\nu}$ are the 
temperatures 
at the epochs of magnetic field generation and 
of neutrino decoupling, respectively.

For the magnetic scalar anisotropic stress trispectrum, ${\cal R}$ in Eq. (4) gets replaced by $\mathcal{R}_p$ and $\zeta_{_{1234}}$ 
becomes $\left[ \zeta_{_{1234}} \right]_{\Pi} = \langle \Pi_B(\kk_{_1})\Pi_B(\kk_{_2})
\Pi_B(\kk_{_3})\Pi_B(\kk_{_4})\rangle$.
The full technical details of the calculation  
of the magnetic scalar anisotropic stress trispectrum will be presented separately \cite{TSS11_passive_trispec}. 
We give below the results considering only the $\ssvec$-independent angular mode-coupling terms for equal-sided configurations. 
In this case
\EQA
\!\left[ \zeta_{_{1234}} \right]_{\Pi} \; 
&\simeq& \;\;\delta(\kk_1 + \kk_2 + \kk_3 + \kk_4) \, \times \, {3^4 \, \xi} \,\times \nonumber \\
&& \!\!\!\!\!\!\!\!\!\!\!\!\!\!\!\!\!\!\!\!\!\!\!\!\!\!\!\!\!\!\!\!
\frac{8 \, (24\pi)\, A^4 \,k_1^{2n+3} k_2^n k_3^n }{(8\pi\rho_0)^4}\left[ \frac{(2^{n/2})(4n+3)-(n+3)}{(4n+3)(n+3)} \right]\!.
\label{passive_zeta}
\ENA
Here, $\xi$ is a configuration-dependent 
number that is the sum of all $\ssvec$-independent angular terms. 
This sum involves terms like $\theta_{ab} = \kkk_a \cdot \kkk_b$ that are constant for a 
given $(\kk_1,\kk_2,\kk_3,\kk_4)$ configuration. Values for $\xi$ range between 2 and 14 
for equal-sided trispectrum configurations: collinear, square, rhombus and tetrahedron. 
We adopt a typical value $\xi \simeq 10$.
This leads to a reduced trispectrum 
\EQA
\left[ T^{l_{_1}l_{_2}}_{l_{_3}l_{_4}} (L)\right]_{\Pi}
&\simeq&  \left( 3~\frac{\mathcal{R}_p}{\mathcal{R}} \right)^4 ~\xi~\left[- T^{l_{_1}l_{_2}}_{l_{_3}l_{_4}} (L)\right]_{\Omega} \nonumber \\
&\simeq& 1.1 \times 10^{-19}  ~\left( \frac{\xi}{10} \right) \left(\frac{n+3}{0.2}\right)^3 \left(\frac{B_{-9}}{3}\right)^8 \nonumber \\
&& \quad \times \frac{ h_{l_1 L \, l_2} \, h_{l_3 L \, l_4}}{l_1(l_1+1) l_2(l_2+1) l_3(l_3+1)}.
\label{trispec_aniso_stress}
\ENA
We have used $T_B \simeq 10^{14}$ GeV (corresponding to the reheating temperature) 
and $T_{\nu} \simeq 10^{-3}$ GeV. 
We see that the amplitude of the trispectrum sourced by $\Pi_B$ for equal-sided quadrilateral 
configurations is approximately $10^{10}$ 
times larger than that sourced by $\Omega_B$. 
Comparison with the trispectrum from inflationary perturbations [Eq. \ref{trispectrum_fNL}] 
gives a magnetic field constraint of 
\EQ
B_0 \lesssim 1.3 \text{ nG},
\label{B_Pi_tau}
\EN
using the positive-sided limit $\tau_{NL} < 33000$ from WMAP5 data \cite{smidt10}. 
This is approximately twice as strong as the 2.4 nG $B_0$ limit obtained from the $\Pi_B$ bispectrum \cite{TSS10} 
and does not assume any particular model of inflation or any relation between $\tau_{NL}$ and $f_{NL}$.
However, for those theories of inflation, which lead to 
$\tau_{NL} = \left( 6/5 \,f_{NL}\right)^2 $ \cite{ByrnesSasakiWands06,NG_inflation},
we could perhaps use the relatively tighter limits for $f_{NL}$.  
To be conservative we take the two-sigma limits $-10 < f^{local}_{NL} < 74$ on the best constrained local $f_{NL}$, 
obtained from searching for the CMB bispectrum signal 
in WMAP7 data \cite{komatsu_wmap7}. This gives primordial magnetic field limits of 
\EQ
B_0 \lesssim 0.7 \text{ nG} \qquad \text{and} \qquad B_0 \lesssim 1.1 \text{ nG},
\label{B_Pi_fNL}
\EN
respectively, for the negative and positive $f^{local}_{NL}$ limits.
If one uses the two-sigma limits for $f^{equil}_{NL}$, then the 0.7 nG limit becomes 0.6 nG 
and for $f^{orthog}_{NL}$ it becomes 1.5 nG. However, the uncertainties $\sigma_{f_{NL}}$ for equilateral and orthogonal 
configurations are 7 and 5 times larger compared to the local configuration \cite{komatsu_wmap7}. 
Staying with the best determined $f^{local}_{NL}$ limits thus results in sub-nG upper limits on $B_0$. 
The expected $\Delta f_{NL} < 5$ \cite{komatsu_spergel01} from Planck data will imply even tighter sub-nG 
magnetic field upper limits from the scalar anisotropic stress trispectrum. 
Future consideration of magnetic vector and tensor modes in the trispectrum 
is likely to give additional constraints on primordial magnetic fields.

In summary, we have calculated for the first time the CMB trispectrum sourced by primordial magnetic fields.
The magnetic energy density trispectrum allows us to place stronger limits on the 
primordial magnetic field compared to a similar calculation with the magnetic energy density bispectrum 
\cite{SS09,caprini09_cai10}.
Further, the trispectrum due to magnetic scalar anisotropic stress leads to the tightest constraint 
on large-scale magnetic fields
of $\sim$ 0.7 nG, approximately 3 times as strong as the corresponding bispectrum limit ($\sim$ 2.4 nG)\cite{TSS10}. 
The trispectrum's sensitivity is illustrated by the magnetic to inflationary trispectrum ratio, which is $\sim 10^3$ 
compared to $\sim 1$ for the bispectrum (taking $f_{NL}\sim100$ and $B_0 \sim 3 \text{ nG}$).
The relative contribution of different configurations to the trispectrum is different for magnetic compared to 
inflationary trispectra and will be useful to distinguish between them.
We also note that the magnetic field limit at Mpc scales derived from only the scalar magnetic CMB trispectrum 
is already better than the limit ($\sim$ 2-6 nG) \cite{yam10_pao10_SL10} from the combined 
scalar, vector and tensor modes in the magnetic CMB power spectrum.
Therefore, the trispectrum turns out to be 
a new and more powerful probe of large-scale primordial magnetic fields.

\textit{Acknowledgements} PT and TRS acknowledge the IUCAA Associateship Program as well as 
the facilities at the IUCAA Resource Center, University of Delhi. PT acknowledges support 
from Sri Venkateswara College, University of Delhi, in pursuing this work. TRS acknowledges support 
from CSIR India via grant-in-aid no. 03(1187)/11/EMR-II. We thank the referees for useful comments.

\end{document}